\pdfoutput=1

\documentclass[12pt]{article}

\usepackage{amssymb}
\usepackage{latexsym}
\usepackage[square,sort,comma,numbers]{natbib}
\usepackage[affil-it]{authblk}
\usepackage{url}
\usepackage{xcolor}
\definecolor{newcolor}{rgb}{.8,.349,.1}

\usepackage{graphicx}
\usepackage{subfigure}
\usepackage{amsmath}
\usepackage{tikz}
\usetikzlibrary{arrows, decorations.pathmorphing, backgrounds, positioning, fit, petri, automata}
\definecolor{yellow1}{rgb}{1,0.8,0.2}

\providecommand{\keywords}[1]
{
  \small	
  \textbf{\textit{Keywords---}} #1
}

\begin{document}

\thispagestyle{empty}

\title{A Fast and Effective Method of Macula Automatic Detection for Retina Images}

\author[2]{Yukang Jiang}
\author[1]{Jianying Pan}
\author[2]{Yanhe Shen}
\author[2]{Jin Zhu}
\author[1]{Jiamin Huang}
\author[1]{Huirui Xie}
\author[3, *]{Xueqin Wang}
\author[1, *]{Yan Luo }

\affil[1]{Country State Key Laboratory of Ophthalmology, Image Reading Center, Zhongshan Ophthalmic Center, Sun Yat-Sen University, Guangzhou, 510275, China}
\affil[2]{Southern China Center for Statistical Science, Department of Statistical Science, School of Mathematics, Sun Yat-Sen University, Guangzhou, 510275, China}
\affil[3]{School of Management, University of Sciences and Technology of China, Hefei, 230026, China}

\affil[*]{Corresponding author}

\maketitle

\begin{abstract}
Retina image processing is one of the crucial and popular topics of medical image processing. The macula fovea is responsible for sharp central vision, which is necessary for human behaviors where visual detail is of primary importance, such as reading, writing, driving, etc. This paper proposes a novel method to locate the macula through a series of morphological processing. On the premise of maintaining high accuracy, our approach is simpler and faster than others. Furthermore, for the hospital's real images, our method is also able to detect the macula robustly.

\end{abstract}


%\linenumbers

%% main text
\section{Introduction}

Diabetic retinopathy (DR) is the leading cause of blindness in the working-age population of the developed world \citep{leontidis2017new, mookiah2013computer}. Moreover, it is expected that the number of people suffered from DR will rise to 191 million by 2030 \citep{kaur2018generalized, shaw2010global, khojasteh2018fundus}. However, vision loss due to DR can be prevented if DR is diagnosed in its early development stages \citep{khojasteh2018fundus}. DR diagnosis requires the detection of macula fovea on the retina, performed by visual examination of eye fundus images. The macula fovea is responsible for the sharp central vision, which is necessary for humans in activities where visual detail is of primary importance, such as reading, writing, driving, etc. Besides, macular degeneration damages human vision irreversibly, so the identification of the macular area is significant.

A large number of image processing techniques and algorithms have been proposed to detect the macula. Most of the existing fovea detection methods include two sequential stages. In the first stage, the optic disc (OD) center is detected, and a region of interest is defined by using the known average distance between the fovea and the OD location. They were separated by a constant distance of 2.5 OD diameter approximately. In the second stage, the fovea location is obtained by exploiting the fovea's visual appearance of in the region extracted in the first stage \citep{Dashtbozorg2016Automatic}. \citet{Li2004Automated} locates the macular according to the OD location and vascular arches. Many methods used the vascular arches and OD to find a region of interest where the fovea location is. \citet{Yu2011Fast} locate the macular by selecting the lowest response of a template matching. In the method introduced by \citet{Gegundez2013Locating}, they detect the macula by defining a region of interest concerning the OD location and the vascular tree. And \citet{Chin2013Automatic} uses the OD location information, arched blood vessels, and vascular density to locate the macula. \citet{Dashtbozorg2016Automatic} presents an automatic OD and fovea detection technique using an innovative super-elliptical filter.

Besides, there are machine learning methods to recognize the macula, which generally consist of three stages: candidate exudate detection, feature extraction, and classification. In the method proposed by \citet{Li2015AUTOMATIC}, the exudate candidate regions using the background subtraction technique and morphological technique and the basic properties of exudates are extracted as features for classification by the support vector machine classifier. The recent paper \citep{Bhosale2017Automatic} about macula detection is also using machine learning to analyze the Age-related Macular Degeneration. 

To improve the detection accuracy, the above methods can be mixed. The method presented by \citet{Aquino2014Establishing} utilizes both the visual and the anatomical features for the fovea detection and improve the obtained fovea center estimation when the fovea is detectable in the image.

However, these methods have some drawbacks, which are time-consuming because they need to combine many different methods. Moreover, most of them detect the macula with the information of OD. So if the OD positioning failed, then the location of the macula must be wrong. \citet{Zheng2014A} proposes a method to detect macular firstly before optic disk location or vessel detection using directional local contrast filter and local vessel density feature. And \citet{Kamble2017Localization} presents an approach for fast and accurate localization of OD and fovea using a one-dimensional scanned intensity profile analysis. 

In the paper, we propose a quite fast method to locate the macula through a series of morphological processing. With a common modern laptop equipped Intel i5 processor and 4GB RAM, our algorithm can accomplish the whole detection procedure within half a second on average, faster than most of the current methods. Our macula detection approach is also powerful except for speediness, achieving a desirable accuracy in the real retina images dataset. Moreover, because the detection process does not involve complicated computation, it does not require high-performance computing power and memory hardware to be accessible to most hospitals. And the flow chart of processing is shown below.

\begin{center}
\begin{tikzpicture} [
squarednode/.style={rectangle, fill=yellow1,draw=none,text=black},
]

\node        (zero)                {Input Image};
\node[squarednode]         (first)    [below=of zero]  {Grayscale Conversion};
\node[squarednode]         (second)   [below=of first]            {Morphological Transformation};
\node[squarednode]          (third)   [below=of second]      {Histogram Equalization};
\node[squarednode]         (forth)  [below=of third]      {Otsu's Method Selection};
\node[squarednode]         (fifth) [below=of forth]        {Final Area Selection};
\node        (sixth) [below=of fifth]        {Output Image};

\draw[->] (zero.south) -- node[right]{} (first.north);
\draw[->] (first.south) -- node[right]{} (second.north);
\draw[->] (second.south) -- node[right]{} (third.north);
\draw[->] (third.south) --  node[right]{} (forth.north);
\draw[->] (forth.south) -- node[right]{}  (fifth.north);
\draw[->] (fifth.south) -- node[right]{}  (sixth.north);

\end{tikzpicture} 
\end{center}

The remaining sections are organized as follows. The proposed algorithm is illustrated in Section \ref{macula-detection-approach}. Next, in Section 3, the algorithm is applied to a color fundus images dataset collected from the hospital to validate its performance. Section \ref{conclusion} provides a short conclusion and discussion.

\section{Macula Detection Approach}\label{macula-detection-approach}

\subsection{Image Preprocessing}
\subsubsection{Grayscale Conversion}

First, all color channels from the fundus image (Red, Green, Blue) can be extracted. The green channel shows the best contrast between the vessels and background, usually used for vessels and veins segmentation. Besides, it contains most of the information about macula. The red channel also shows macula and less information about veins and vessels. Therefore, in most articles, the red or green channel of the fundus image is used.

\begin{figure} [!t]
	\centering
	\subfigure[Original image]{
		\begin{minipage}{20mm}
			\centering
  			\includegraphics[width=20mm]{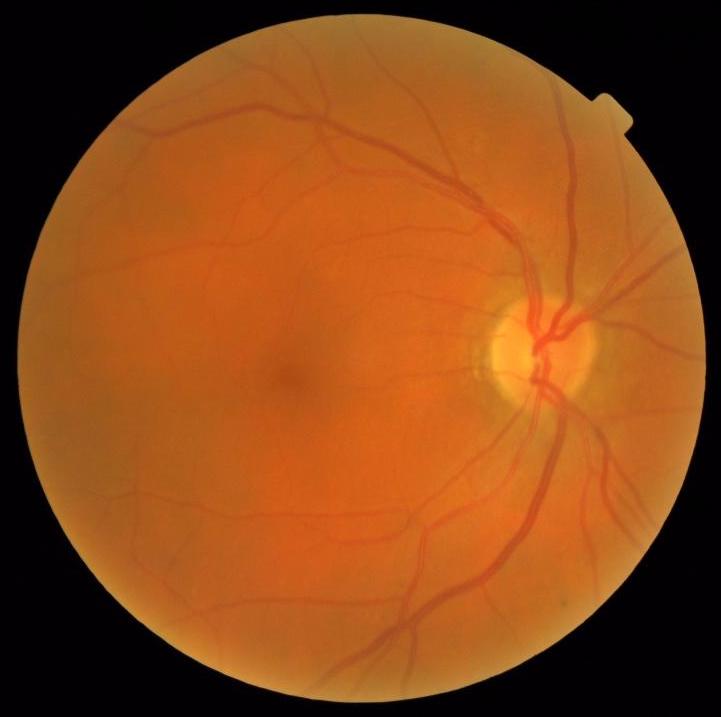}
		\end{minipage} 
	}\subfigure[Red channel]{
		\begin{minipage}{20mm}
			\centering
  			\includegraphics[width=20mm]{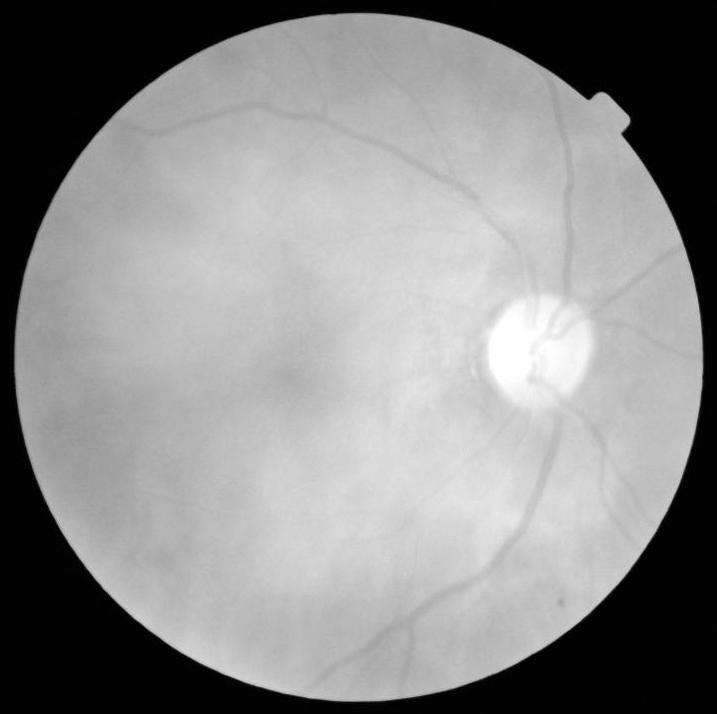}
		\end{minipage} 
	}\subfigure[Green channel]{
		\begin{minipage}{20mm}
			\centering
  			\includegraphics[width=20mm]{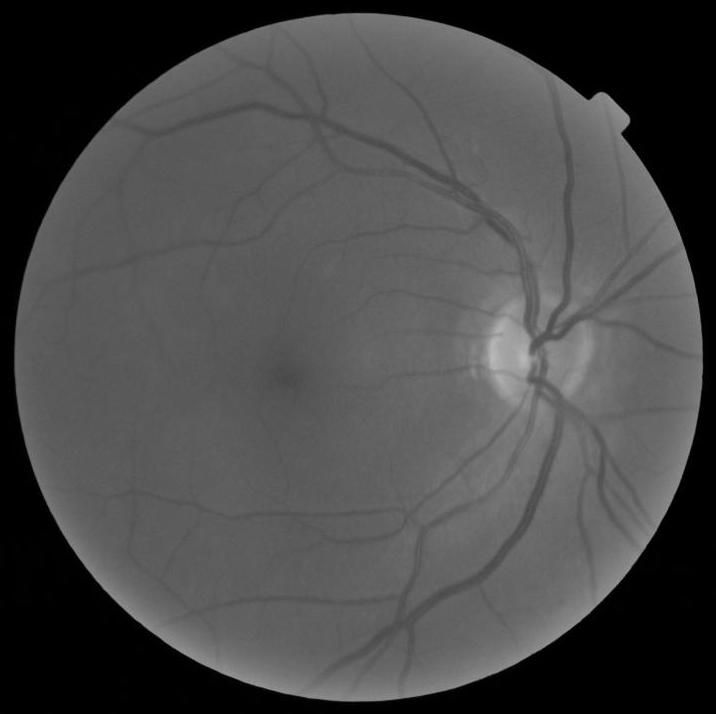}
		\end{minipage} 
	}\subfigure[Blue channel]{
		\begin{minipage}{20mm}
			\centering
  			\includegraphics[width=20mm]{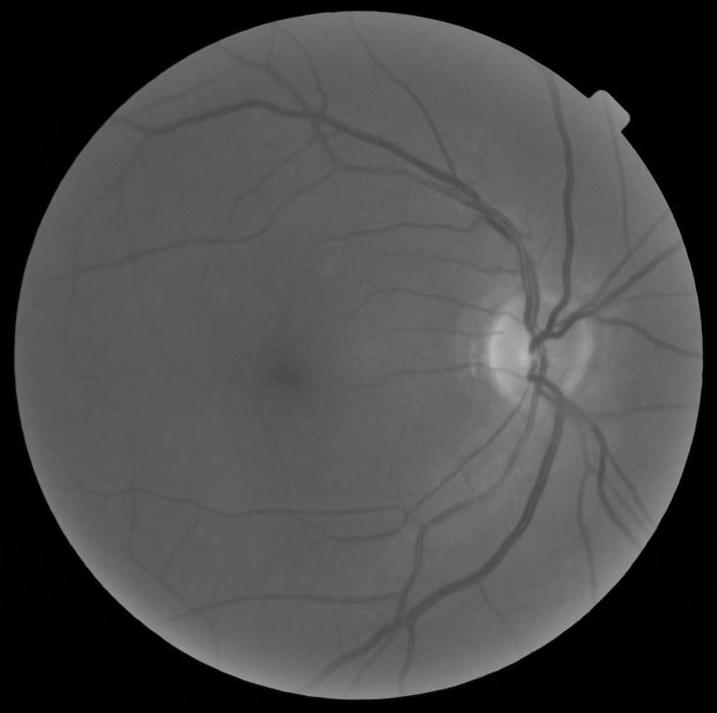}
		\end{minipage} 
	}
	\caption{The original image and RGB channels of fundus retinal image.}\label{fig:4pic}
\end{figure}

In fact, from figure 1, we conclude that each layer contains the macular information more or less. And we can see the location of the macula from the original image. So we combine the three layers by converting the RGB image to a grayscale image which we will be based on for the follow-up processing. The formula represented as:
\begin{equation}
\mathrm{Grayscale} = 0.299 \times \text{Red} + 0.587 \times \text{Green} + 0.114 \times \text{Blue},
\end{equation}
Grayscale can be regarded as the quantification of luminance, and RGB is defined as three-wavelength values. When converting, it needs to consider the sensitivity curve of different wavelengths for human eyes. And the coefficients corresponding to the three RGB layers are as shown in (1).

In the following, we will utilize two pictures as an example to describe the details of the macula detection from the beginning to the end.

\begin{figure} [!t]
	\centering
	\subfigure[Image with optic disc.]{
		\begin{minipage}{40mm}
			\centering
  			\includegraphics[width=40mm]{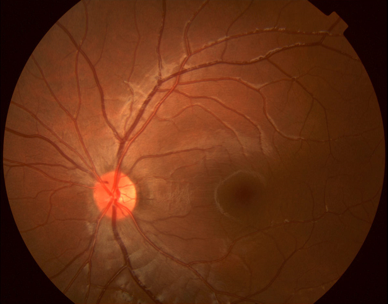}
		\end{minipage} 
	}\subfigure[Image without optic disc.]{
		\begin{minipage}{40mm}
			\centering
  			\includegraphics[width=40mm]{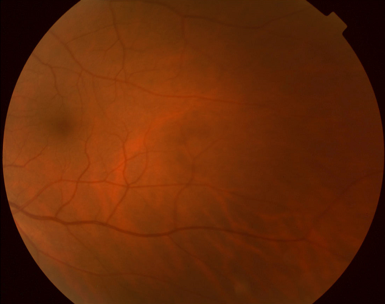}
		\end{minipage} 
	}
	\subfigure[Grayscale image with optic disc.]{
		\begin{minipage}{40mm}
			\centering
  			\includegraphics[width=40mm]{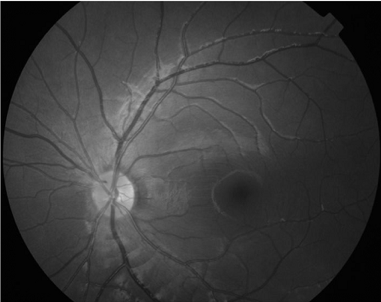}
		\end{minipage} 
	}\subfigure[Grayscale image without optic disc.]{
		\begin{minipage}{40mm}
			\centering
  			\includegraphics[width=40mm]{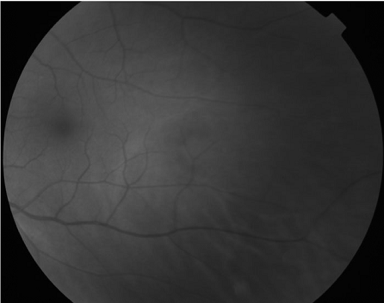}
		\end{minipage} 
	}
	\caption{Original retina images and their grayscale images throuth combining the three layers.}\label{fig:4pic}
\end{figure}

\subsubsection{Morphological Transformation}
Morphology is a broad set of image processing operations that process images based on shapes. In most articles about macula detection, the morphological transformation is an indispensable operation, and this article is no exception.

Grayscale structuring elements are also functions of the same format, called "structuring functions". Let $x$ and $y$ represent the coordinates of each pixel. Denoting an image by $f(x)$ and the structuring function by $b(x)$, the grayscale dilation of $f$ by $b$ is given by
\begin{equation}
(f\oplus b)(x)=\sup _{{y\in E}}[f(y)+b(x-y)],
\end{equation}
where ``sup'' denotes the supremum.

Similarly, the erosion of $f$ by $b$ is given by
\begin{equation}
(f\ominus b)(x)=\inf _{{y\in E}}[f(y)-b(y-x)],
\end{equation}
where ``inf'' denotes the infimum.

Just like in binary morphology, the opening and closing are given respectively by
\begin{equation}
f\circ b=(f\ominus b)\oplus b,
\end{equation}
and
\begin{equation}
f\bullet b=(f\oplus b)\ominus b.
\end{equation}

Dilation expands an image and erosion shrinks it. Opening generally smooths the contour of an object, breaks narrow isthmuses, and eliminates thin protrusions. The closing also tends to smooth sections of contours, but, as opposed to opening, it generally fuses narrow breaks and long thin gulfs, eliminates small holes, and fills gaps in the contour \citep{Gonzalez2010Digital}.

Moreover, the top-hat transform of $f$ is given by: 
\begin{equation}
T_{top}(f, b)=f-f\circ b,
\end{equation}
 where $\circ$  denotes the opening operation.

This transformation is useful for enhancing the detail in the presence of shading. And the bottom-hat transform of $f$ is given by:
\begin{equation}
T_{bottom}(f, b)=f\bullet b-f,
\end{equation}
where $\bullet$  is the closing operation.

\begin{figure} [!t]
	\centering
	\subfigure[Image with optic disc.]{
		\begin{minipage}{40mm}
			\centering
  			\includegraphics[width=40mm]{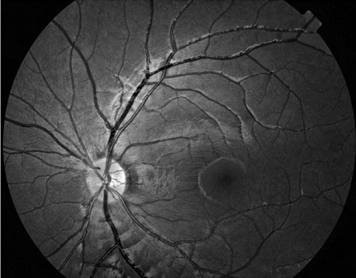}
		\end{minipage} 
	}\subfigure[Image without optic disc.]{
		\begin{minipage}{40mm}
			\centering
  			\includegraphics[width=40mm]{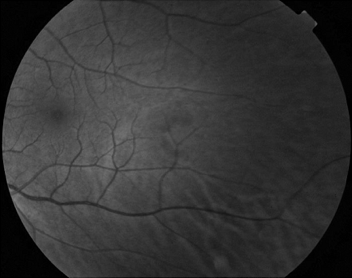}
		\end{minipage} 
	}
	\caption{The two defferent images containing optic disc or not after morphological transformation.}\label{fig:morphological}
\end{figure}

In the actual image processing, we use the detailed algorithm as follows: 
\begin{equation}
Image + T_{top}(Image, Disk15) - T_{bottom}(Image, Disk15),
\end{equation}
$Disk15$ means a disk-shaped structuring element, which radius is 15 pixels. 
 And after processing, the two images are visualized in Figure \ref{fig:morphological}.

\subsubsection{Adaptive Histogram Equalization}
\begin{figure} [!t]
	\centering
	\subfigure[Image with optic disc.]{
		\begin{minipage}{40mm}
			\centering
  			\includegraphics[width=40mm]{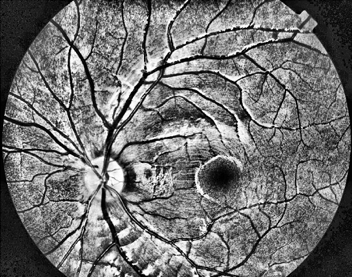}
		\end{minipage} {}
	}\subfigure[Image without optic disc.]{
		\begin{minipage}{40mm}
			\centering
  			\includegraphics[width=40mm]{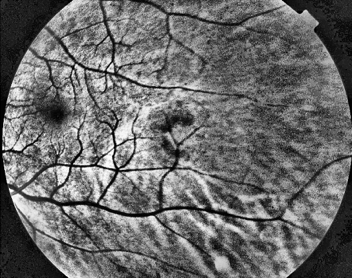}
		\end{minipage} 
	}
	\caption{The images after histogram equalization to highlight the macula and blood vessels.}\label{fig:histogram}
\end{figure}

To further enhance the macular area's detail, we use Adaptive Histogram Equalization (AHE) to process the image. AHE is applied for contrast enhancement. And dark regions, including vessels, macular areas, and so on, are dominant after contrast enhancement. It usually increases the global contrast of images. By this adjustment, the intensities can be better distributed on the histogram. This allows for areas of lower local contrast to gain a higher contrast. Histogram equalization accomplishes this by effectively spreading out the most frequent intensity values. The first step is to count the number of occurrences of each gray level in the gray histogram. The second step is to accumulate normalized gray histograms. And the final step is to calculate new pixel values. The two images after adaptive histogram equalization are demonstrated in Figure \ref{fig:histogram}.

For eliminating the useless noise, we also need to erode and dilate the images.
\begin{figure} [!t]
	\centering
	\subfigure[Image with optic disc.]{
		\begin{minipage}{40mm}
			\centering
  			\includegraphics[width=40mm]{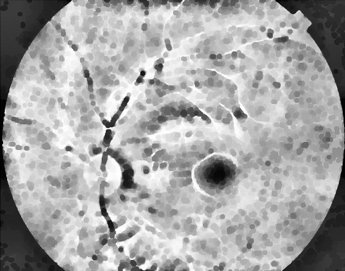}
		\end{minipage} 
	}\subfigure[Image without optic disc.]{
		\begin{minipage}{40mm}
			\centering
  			\includegraphics[width=40mm]{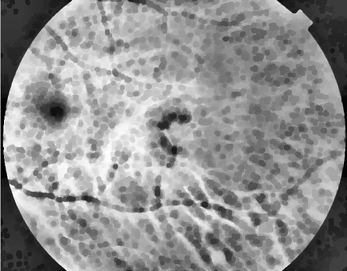}
		\end{minipage} 
	}
	\caption{The images use the top-hat as well as bottom-hat transform to eliminate blood vessels and retain the macula.}\label{fig:4pic}
\end{figure}

\subsection{Otsu's method}
Next, we aim to find the macular area and choose Otsu's method (the maximum between-class variance method), which uses the idea of clustering. Through this method, the number of grayscale images is divided into two parts by grayscale, calculated by variance to find a suitable gray level to divide. So we can utilize Otsu algorithm in the binarization to select the threshold for binarization automatically. Besides, Otsu algorithm is considered the best algorithm for threshold selection in image segmentation, and the calculation is simple and not affected by image brightness and contrast.

The algorithm assumes that the image contains two classes of pixels following bi-modal histogram (foreground pixels and background pixels). It then calculates the optimum threshold separating the two classes so that their combined spread (intra-class variance) is minimal, or equivalently (because the sum of pairwise squared distances is constant) so that their inter-class variance is maximal. \citep{otsu1979a}

Assuming that the gray level of the original image is $M$ and the total number of pixels is $N$. The number of pixels with grayscale $i$ is $n_i$. 

First, we normalize the gray level
\begin{equation*}
P_i = \frac{n_i}{N},
\end{equation*}
and set the threshold of segmentation as $t$. The grayscale is divided into two categories, the probability of each category is
\begin{equation*}
\omega _{0}(t) =\sum _{i=0}^{t-1}P_i, \ \ \ \ \ \omega _{1}(t)=\sum _{i=t}^{M-1}P_i,
\end{equation*}
where $\omega _{0}(t) + \omega _{1}(t) = 1 $.

The average grayscale appearance of each category is
\begin{equation*}
\mu _{0}(t)=\sum _{i=0}^{t-1}i{\frac {P_i}{\omega _{0}(t)}},  \ \ \ \mu _{1}(t)=\sum _{i=t}^{M-1}i{\frac {P_i}{\omega _{1}(t)}},
\end{equation*}
and the overall average grayscale is
\begin{equation*}
\mu=\sum _{i=0}^{M-1}iP_i.
\end{equation*}

Then, we exhaustively search for the threshold that maximizes the intra-class variance, defined as a weighted sum of variances of the two classes:
\begin{equation}
\begin{aligned}\sigma^{2}(t)&=\omega _{0}(\mu _{0}-\mu)^{2}+\omega _{1}(\mu _{1}-\mu)^{2}\\&=\omega _{0}(t)\omega _{1}(t)\left[\mu _{0}(t)-\mu _{1}(t)\right]^{2}\end{aligned}
\end{equation}

Finally, we step through all possible thresholds $t=1,\ldots$, and maximum the intra-class variance $\sigma^{2}(t)$. Due to the particularity of the macula. We subtract a threshold value of 0.2 from the final value $t$ as a threshold value for better results. If the threshold value is larger than 1, we take 1 (white), and if it is less than 0, we take 0 (black). Figure 6 shows the results after Otsu's method.

\begin{figure} [!t]
	\centering
	\subfigure[Image with optic disc.]{
		\begin{minipage}{40mm}
			\centering
  			\includegraphics[width=40mm]{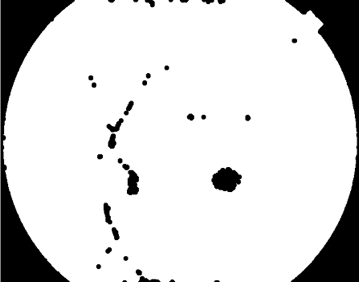}
		\end{minipage} 
	}\subfigure[Image without optic disc.]{
		\begin{minipage}{40mm}
			\centering
  			\includegraphics[width=40mm]{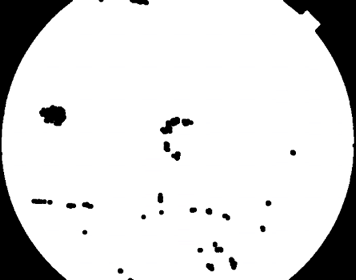}
		\end{minipage} 
	}
	\caption{Image were processed by Otsu's method for segment the images.}\label{fig:4pic}
\end{figure}

\subsection{Area Selection}

We select the final region by calculating the area of each connected domain and retain the area that more than 400 and less than 5000. Because most of the macular areas are circular, we judge the location of the macula by whether it is a circle or not through the formula
\begin{equation}
T = \frac{4A}{\pi  P^2},
\end{equation}
where $A$ is the connected domain area, $P$ is the longest length of the connected domain. The value of $T$ is between 0 and 1, and the closer to 1, the more circular the connected domain is. And the connected domain is circular if and only if $T = 1$. 

Finally, we filter out the largest connected domain and determine the center of the connected domain, the center of the foveal location. At the same time, to improve the recognition accuracy, for the images that are easy to identify the OD, we can determine the location of the macula with the information of OD.

\begin{figure} [!t]
	\centering
	\subfigure[Image with optic disc.]{
		\begin{minipage}{40mm}
			\centering
  			\includegraphics[width=40mm]{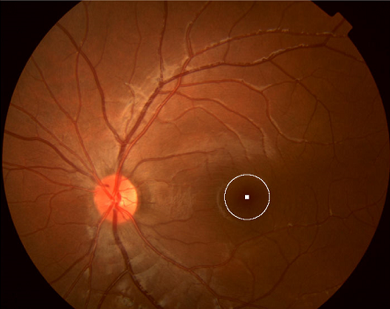}
		\end{minipage} 
	}\subfigure[Image without optic disc.]{
		\begin{minipage}{40mm}
			\centering
  			\includegraphics[width=40mm]{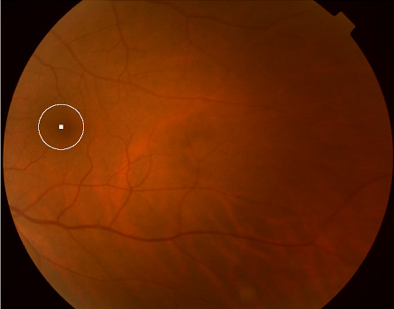}
		\end{minipage} 
	}
	\caption{The results of detection locating the macula accurately.}\label{fig:4pic}
\end{figure}

\section{Real Data}
This section employed our proposed algorithm on 254 real retina images collected by Image Reading Center Zhongshan Ophthalmic Center of Sun Yat-sen University. The data set contained 247 images, including the macula. The image was stretched by $700 \times 1050$. 

\begin{figure} [!t]
	\centering
  			\includegraphics[width=80mm]{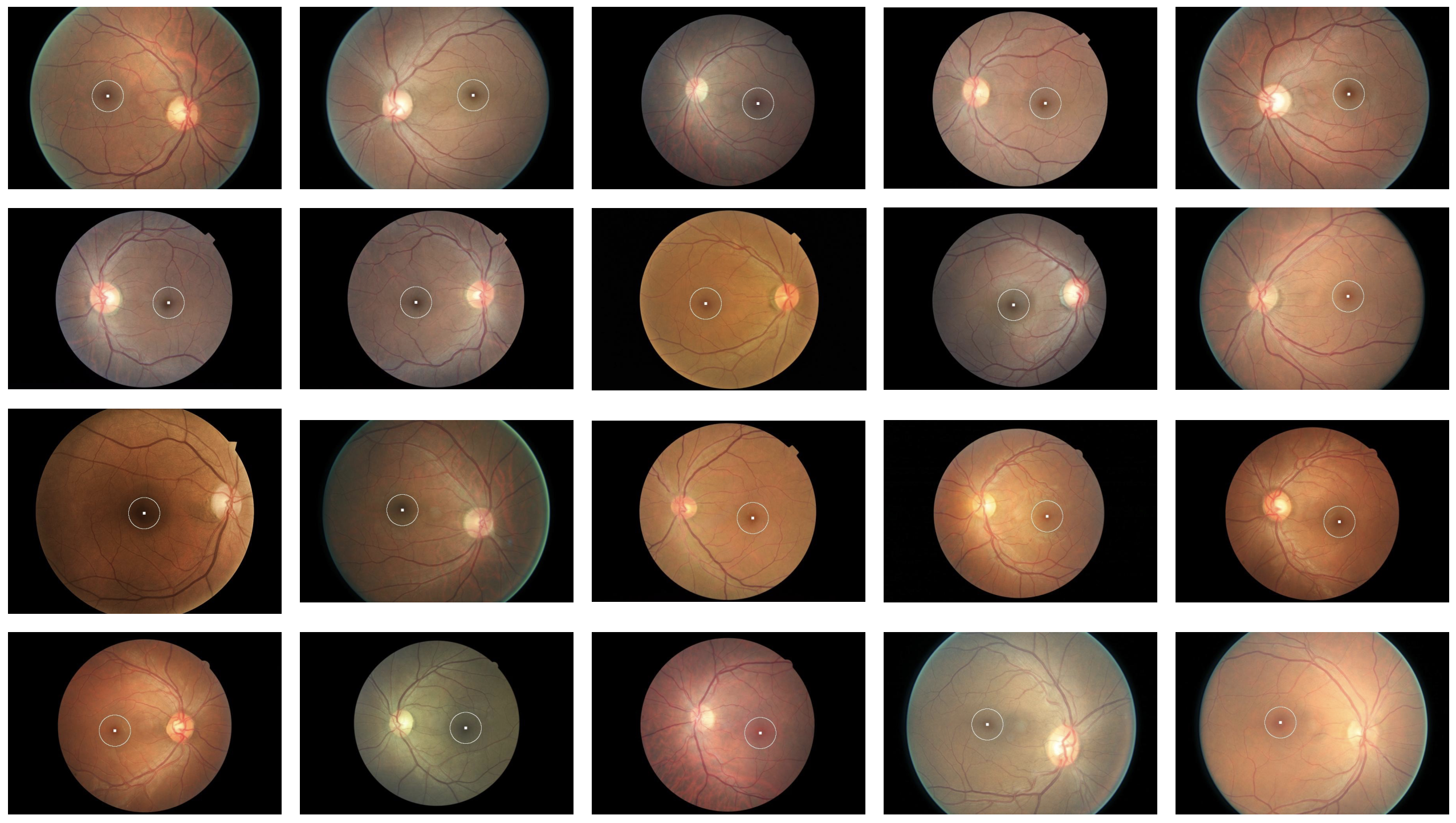}
	\caption{Final results of detection.}\label{real-detection}
\end{figure}

A part of typical macula detection results based on our proposed algorithm is demonstrated in Figure \ref{real-detection}. And it shows that our method can successfully detect the location of the macula.

Furthermore, we discuss the classification accuracy of our method. Our detection results are evaluated by 2 experienced image readers. Besides, the recognition accuracy is shown in Table \ref{classification-accuracy}.

\begin{table}[!t]
	\center
	\begin{tabular}{ccc}
		\hline 
		 & Actual 0 & Actual 1 \\ 
		\hline 
		Predicted 0 & 7 & 8 \\ 
		Predicted 1 & 0 & 239 \\ 
		\hline 
	\end{tabular}
	\caption{Recognition Accuracy}
\end{table}\label{classification-accuracy}

Table \ref{classification-accuracy} shows that if we identify a macula in a retina image, it must be a macula. And it means that the false positive rate is 0\%. The sensitivity is 96.8\%, and the specificity is 100\%.

\section{Conclusion}\label{conclusion}
This paper proposes a simple method to locate the macula through a series of morphological processing. On the premise of maintaining high accuracy, our approach is simpler and faster than others and does not use the blood vessel, optic disc, and other location information. For the real images of the hospital, our method can also detect the macula automatically and robustly. 

\newpage

\bibliographystyle{plainnat}
\bibliography{refs}

\end{document}